\documentclass[conference]{IEEEtran}

\usepackage{amsmath,amssymb}

\usepackage{url,xspace}
\usepackage[english]{babel}
\usepackage{xcolor}
\usepackage{graphicx}
\usepackage{booktabs}

\usepackage{hyperref}

 \newlength\figureheight 
  \newlength\figurewidth 

\newcommand{\refeq}[1]{Eq.(\ref{#1})}

\newcommand{\picwidth}{\columnwidth}

\hypersetup{
    colorlinks,
    linkcolor={red!50!black},
    citecolor={blue!50!black},
    urlcolor={blue!80!black}
}

\graphicspath{{figs/}}

\newcommand{\tme}{\ensuremath{\%\mathrm{TME}}\xspace}
\newcommand{\pow}{\ensuremath{\%\mathrm{PoW}}\xspace}
\newcommand{\gob}{\ensuremath{\%\mathrm{GoB}}\xspace}

\newcommand{\tmen}{\ensuremath{\mathrm{TME}}\xspace}
\newcommand{\pown}{\ensuremath{\mathrm{PoW}}\xspace}
\newcommand{\gobn}{\ensuremath{\mathrm{GoB}}\xspace}

\newcommand{\eqv}[2]{#2\ensuremath{\triangleq}'#1'}

\title{Formal Definition of QoE Metrics}

\newcommand{\mail}[1]{Email: \href{mailto:#1}{#1}}

\author{\IEEEauthorblockN{Tobias Ho\ss{}feld\IEEEauthorrefmark{1}, Poul E. Heegaard\IEEEauthorrefmark{2}, Mart{\'\i}n Varela\IEEEauthorrefmark{3}, Sebastian M\"oller\IEEEauthorrefmark{4}} 
\IEEEauthorblockA{\IEEEauthorrefmark{1}Modeling of Adaptive Systems, University of Duisburg-Essen, Germany, \mail{tobias.hossfeld@uni-due.de}} 
\IEEEauthorblockA{\IEEEauthorrefmark{2} Norwegian University of Science and Technology (NTNU), \mail{poul.heegaard@item.ntnu.no}} 
\IEEEauthorblockA{\IEEEauthorrefmark{3} VTT Technical Research Centre of Finland --- Communication Systems, \mail{Martin.Varela@vtt.fi}}
\IEEEauthorblockA{\IEEEauthorrefmark{4} Quality and Usability Lab, TU Berlin, Germany, \mail{Sebastian.Moeller@telekom.de}}}

\begin{document}

\maketitle
\thispagestyle{plain}
\pagestyle{plain}
\begin{abstract}
This technical report formally defines the QoE metrics which are introduced and discussed in the article ``QoE Beyond the MOS: An In-Depth Look at QoE via Better Metrics and their Relation to MOS'' by Tobias Ho\ss{}feld, Poul E. Heegaard, Mart{\'\i}n Varela, Sebastian M\"oller, accepted for publication in the Springer journal 'Quality and User Experience' \cite{hossfeld2016qoeBeyond}.
Matlab scripts for computing the QoE metrics for given data sets are available in GitHub \cite{hossfeld2016github}. 
\end{abstract}

\section{Definition of QoE Metrics}

\newcommand{\mysection}[1]{\subsection{#1}}

\mysection{Preamble}
We consider studies where users are asked their opinion on the overall quality (QoE) of a specific service.
The {\em subjects} (the participants in a study that represent users), rate the quality as a {\em quality rating} on a {\em quality rating scale}. 
As a result, we obtain an {\em opinion score} by interpreting the results on the rating scale numerically. 
An example is a discrete 5-point scale with the categories \eqv{bad}{1}, \eqv{poor}{2}, \eqv{fair}{3}, \eqv{good}{4}, and
\eqv{excellent}{5}, referred to as an Absolute Category Rating (ACR) scale \cite{Moeller00}. 

Let $U$ be a random variable (RV) that represents the quality ratings, $U \in \Omega$, where $\Omega$ is the rating scale, 
which is also the state space of the random variable $U$. 
The RV $U$ can be either discrete, with probability mass function $f_u$, or continuous, with probability density function $f(u)$.

The notation used in the paper is summarized in Table~\ref{appendix:tab:notation}.
\begin{table}[b!!!!!!!!!!!!!!!]%
\caption{Key variables and notations used in the paper.}
\label{appendix:tab:notation}
\begin{tabular}{p{0.15\columnwidth}p{0.75\columnwidth}}
\toprule
notation & meaning \\
\midrule
MOS & Mean Opinion Score \\
SOS & Standard deviation of Opinion Score \\
ACR & Absolute Category Rating \\
$U$ & random variable (RV) for quality ratings \\
$U^{+}$ & upper limit of a rating scale \\
$U^{-}$ & lower limit of a rating scale \\
$\mathcal{U}$ & set of ratings $\{U_i\}$ for a particular test condition \\
$R$ & number of subjects (= ratings) per test condition \\ 
$J$ & number of tests\\
$\mathbb{S}$ & Statistical definition set ($\{ \Omega, \mathcal{C}, \Sigma , \mathcal{S} \}$)\\
$\Omega$ & quality rating scale  
(same as the sample space of the RV $U$)\\
$\mathcal{C}$ & set of test conditions  \\
$\Sigma$ & set of statistics \\
$\mathcal{S}$ & set of (sufficient) observators \\
$u$ & quality rating value for a test condition (e.g. MOS on the ACR scale)\\
$f_u$ & probability that the quality rating is $u\in \Omega$\\
$\hat{f}_u $ & estimate of $f_u$, i.e. ratio of subjects who rate the test condition with $u$\\ 
S(u) & SOS as a function of mean opinion score $u$\\
$a$ & SOS parameter of SOS hypothesis in Eq.~\eqref{appendix:eq:SOS}\\
$\mathbb{A}_\theta$ & $\alpha$-acceptability, i.e. the probability that ratings are above $\theta$, $P(U>\theta)$ \\
\pow & Poor-or-Worse (in \%) \\
\gob & Good-or-Better (in \%) \\
\tme & Terminate Early (in \%) \\
\bottomrule
\end{tabular}
\end{table}

\mysection{Observations in a study are samples of \texorpdfstring{$U$}{U}}\label{appendix:subsubsect:samples}

In a study, each observation can be regarded as samples $U_{i,r,c}$ of $U$.
Each sample corresponds to the observation $i$ of subject $r$, under test condition $c$, using the rating scale $\Omega$.
Typically, the notation can be simplified, e.g., if the identity of a subject is not an issue, 
and we look at only one test with the same test conditions $c$, then the set of observations 
will form a set of samples $\mathcal{U} = \{ U_i \}$, $i=1,\cdots ,R$.
More details about test conditions are found in the next section.

\mysection{Specification of test conditions}
A study typically specifies a set of test conditions, $\mathcal{C}$, which includes both technical and non-technical factors. 
The set of test conditions contains an invariant (static) part that is unchanged over all tests, $\mathcal{C}_0$, 
and a variable (dynamic) part that changes between each of the $J$ tests, $\mathcal{C}_j$, ($j=1,\cdots ,J$). 
For each test, the test conditions might change according to a pattern specified on a set (of size $k$), $\mathcal{C}_j = \{ c_{j,1}, \cdots , c_{j,k} \}$.  
If $\mathcal{C}_j$ is not the same for all $R$ subjects in test $j$, 
then the notation must be extended and indexed with, $r$, i.e. $\mathcal{C}_{j,r}$.
Furthermore, in some studies we don't have the same number of subjects per test condition.
To specify this, you should add an index to the number of subjects $R_j$ of test $j$.

As an example, consider a study of the effect of a specific sequence of changes in the video quality classes, 
where each subject is expected to rate the overall technical quality of the video.  
In each test a specific set of quality classes, $q_n$, is defined, let's say $N$ classes, 
in an unordered set $\mathcal{Q} = \{ q_1, \cdots , q_N \}$. 
The variable part of $\mathcal{C}$ is then defined as an ordered set $\mathcal{C}_{j} = (c_{j,1}, \cdots, c_{j,k})$, 
where $c$ contains both the quality class, $q$, and the time, $t$, for the quality class changes, $c = (t, q)$.
If the sequence of changes is not the same for each subject in a test, either with respect to $q$ or $t$, 
then the notation must capture this by extending and indexing the $\mathcal{C}_{j}$ with $r$ as described above.
If the test is about the effect of a number of changes, $k$, and not a specific sequence of such, 
then $\mathcal{C}_{j} = (k, \mathcal{Q}$).
Finally, if each test only has one condition, then $\mathcal{C}_{j} = c_{j}$, where $k=1$ and $c_j \in \mathcal{Q}$.

See Table~\ref{appendix:tab:pattern} for a summary of the different options in this example.

\begin{table}[t]%
\centering
\caption{Different examples of test condition sets for the case with the changes in the video quality classes. 
}
\label{appendix:tab:pattern}
\begin{tabular}{p{0.42\columnwidth}p{0.50\columnwidth}}
  \toprule
	Test conditions & Description of test condition for test $j$\\
	\midrule
	$C_j = \{ q_j \}$ & Fixed quality level ($k=1$) \\
	$C_j = \{ k, \mathcal{Q}\}$ & $k$ random changes of quality level \\
	$C_j = \{ (t_1,q_1), \cdots ,(t_k,q_k) \} $ & $k$ specific changes in quality level \\
	$C_{j,r}$ & specific changes for each subject $r$\\
	\bottomrule
\end{tabular}
\end{table}

To give an illustration of how to use this in a real study, let's consider the example of web QoE \cite{hossfeld2016qoeBeyond}.
Here we have~$R=72$ subjects who were downloading~$k=40$ web pages, 
with different content~$w_i$, and page loading time (PLT), $q_i$, which gives $c_j = (w_j,q_j)$. 
The test conditions can be specified either as a sequence of $k=40$ content and impairments, with 
$C_1 = \{ (w_1,q_1), \cdots , (w_{40},q_{40}) \}$, or alternatively, if the specific sequence doesn't matter, 
then the test can be described by $J=40$ separate tests with one test condition each, 
$C_j = \{(w_j,q_j)\}$, $j=1, \cdots , 40$.
In this study, all subjects, $r$, had the same $\mathcal{C}$.

\mysection{\texorpdfstring{$n$}{n}-tuple statistical definition set}
An experimental design of a study needs to specify both technical and non-technical factors, 
and the kind of statistics that will be obtained. 
To make sure that all necessary details about the statistics are included, it can be useful to have a compact notation for this. 
In this paper we propose an $n$-tuple, $\mathbb{S} = \{ \Omega, \mathcal{C}, \Sigma , \mathcal{S} \}$ for this purpose, where
\begin{itemize}
\item $\Omega$ is the {\em rating scale}, i.e. the sample set, $U_i \in \Omega$, e.g. $\{ 1,\cdots , 5\}$, e.g. $[0;6]$.
\item $\mathcal{C}$ is the set of test conditions\footnote{the static part $\mathcal{C}_0$ can be defined outside the $n$-tuple}.
\item $\Sigma$ is the statistics that will be obtained, e.g. MOS, SOS, median, quantiles, etc. 
\item $\mathcal{S}$ is the sufficient observators for the estimators of the statistics in $\Sigma$, e.g. $\sum U_i$, $\sum U_i^2$, etc.
\end{itemize}

The notation here should be considered as an example of a minimal $n$-tuple. 
Necessary extensions might apply.

\mysection{Expected value and its estimate: MOS}
The expected value (sometime referred to as the {\em mean}) of the random variable, is 
\begin{equation}
E[U] = \left\{
   \begin{array}{ll}
     \sum_{u=U^{-}}^{U^{+}} u  f_u & : U\mbox{ is discrete}\\
     \int_{u=U^{-}}^{U^{+}} u  f(u) du & : U\mbox{ is continuous}
   \end{array}
 \right.
\end{equation}
This is an example of a $1$st-order statistics, along with the {\em median} and the {\em mode}. 

The Mean Opinion Score (MOS) is an estimate $\hat{U}$ of $E[U]$.
\begin{equation}\label{appendix:eq:MOS1}
\mbox{u = MOS} = \hat{U} = \sum_{v=1}^{N} v \hat{f}_v 
\end{equation}
where $\hat{f}_u$ is the estimated probability of opinion score $u$, 
\begin{equation}\label{appendix:eq:p}
\hat{f}_u = \frac{1}{R} \sum_{i=1}^{R} \delta_{U_i,u}
\end{equation}
with the Kronecker delta $\delta_{i,j}=1$ if $i=j$ and 0 otherwise.  Substituting $\hat{f}_u$ from~\eqref{appendix:eq:p} in~\eqref{appendix:eq:MOS1} gives, 

\begin{equation}
\hat{U} = \sum_{v=1}^{N} v \hat{f}_v = \sum_{v=1}^{N} v \frac{1}{R} \sum_{i=1}^{R} \delta_{U_i,v} 
\end{equation}

Now, let $U_i = \sum_{v=0}^{N} v \delta_{U_i,v}$, and by reorder the sums, then 
\begin{equation}
\hat{U} =  \frac{1}{R} \sum_{i=1}^{R} \sum_{v=1}^{N} v \delta_{U_i,v} = \frac{1}{R} \sum_{i=1}^{R} U_i
\end{equation}

Hence, we have two, unbiased estimators of $E[U]$,
\begin{equation}
\hat{U} = 
\sum_{v=1}^{N} v \hat{f}_v = \frac{1}{R} \sum_{i=1}^{R} U_i
\end{equation}

\mysection{Standard deviation and its estimate: SOS }
The variance $\sigma_U^2$ of $U$ is
\begin{equation}
Var[U]= \left\{
   \begin{array}{ll}
     \sum_{u=U^{-}}^{U^{+}} (u - E[U])^2  f_u & : U\mbox{ is discrete}\\
     \int_{u=U^{-}}^{U^{+}} (u - E[U])^2 f(u) du & : U\mbox{ is continuous}
   \end{array}
 \right.
\end{equation}
The standard deviation of $U$ is $\sigma_U = \sqrt{Var[U]}$.
This is an example of a $2$nd-order statistics, which contains more information than the mean, since it gives a measure of the uncertainty of the mean. 

The standard deviation of the opinion score (SOS), is an estimate of the $\sigma_U$:
\begin{equation}
S = \sqrt{\frac{1}{R-1} \sum_{i=1}^{R} U_i^2  - \frac{R}{R-1} \hat{U}^2}
\end{equation}
The SOS is estimated over the sample set $\mathcal{U} =\{ U_i \}$ taken from either a discrete or continuous distribution.

It is important to note that the $S$ is the {\em estimated standard deviation of the opinion score}, $U$, which converges to $S \rightarrow \sigma_U$ for a large number of samples, $R$.  This must not be confused with the {\em standard error} of the $\hat{U}$, which is $\mbox{s.e.}(\hat{U}) = S/\sqrt{R}$ and converges to $S/\sqrt{R} \rightarrow 0$ for a large $R$.  

\mysection{SOS as function of MOS}
In~\cite{Hossfeld2011_SOS}, the minimum, $S^{-}(u)$, and the maximum SOS, $S^{+}(u)$ were obtained, as a function of the MOS=$u$.
The minimum SOS is $S^{-}(u) = 0$ on a continuous scale, $[U^{-};U^{+}]$, and 
\begin{equation}
S^{-}(u) = \sqrt{u (2 \lfloor u\rfloor +1)-\lfloor u\rfloor  (\lfloor u\rfloor +1)-u^2} \label{appendix:eq:minSOS}
\end{equation} 
on a discrete scale, $\{U^{-} \cdots U^{+}\}$. 

The maximum SOS is, on both continuous and discrete scales (the scales as above),
\begin{equation}
S^+(u)=\sqrt{-u^2+(U^{-} + U^+)u - U^{-} \cdot U^+}
\end{equation} 

The SOS hypothesis~\cite{Hossfeld2011_SOS}, formulates a generic relationship between MOS and SOS values independent of the type of service or application under consideration. 
\begin{equation}
S(u) = \sqrt{a} \cdot S^{+}(u)
\label{appendix:eq:SOS}
\end{equation} 

It has to be noted that the SOS parameter $a$ is scale invariant when linearly transforming the user ratings and computing MOS and SOS values for the transformed ratings. The SOS parameter allows to compare user ratings across various rating scales. Thus, any linear transformation of the user ratings does not affect the SOS parameter $a$ which is formally proven in Section~\ref{appendix:sec:invarianceSOS}. 
However, it has to be clearly noted that if the participants are exposed to different scales, then different SOS parameters may be observed. 
The parameter $a \in [0;1]$, 
depends on the application or service, and the test conditions, and is derived from subjective tests. 
The computation of the SOS parameter is provided \refeq{eq:computeSOS} in Section~\ref{sec:computationSOS}.

\mysection{Quantiles and Distribution of User Ratings}
The state space $\Omega$ of the random variable $U$ can be either discrete, 
with probability mass function $f_u$, or continuous, with probability density function $f(u)$.  
If an observed set $\mathcal{U} = \{ U_i \}$ from a study contains a large number of observations, 
then the $f_u$ or $f(u)$ can be estimated (the most well know probability density estimation approach is the histogram).  
For most studies in the QoE domain, estimating {\em quantiles} is an alternative option that can be quite useful. 

The $n$'th $q$-quantile of a random variable $U$ is a value $Q_{n/q}$, such that the probability 
$P(U \le Q_n/q) \le n/q$  and $P(U \ge Q_n/q) \le (q-n)/q$, ($0<n<q$).   Well-know quantiles are 
\begin{itemize}
\item $q=2$ : this is called the {\em median} 
\item $q=4$ : this is called the {\em quartile}
\item $q=100$ : this is called the {\em percentile}
\end{itemize}
Assume that we have an ordered set of $R$ quality ratings $\mathcal{U} = \{ U_i \}$ from a study. 
The $q$-quantiles for the ratings can then be estimated by 
\begin{equation}
\hat{Q}_{n/q} = \mathcal{U}^{(h)}
\end{equation} 
where $h= \lceil R n/q \rceil$, ($0<n<q$)\footnote{other estimators for $\hat{Q}_q $ exists}.

As shown above, this is also called the $\alpha$'th percentiles, with $\alpha = 10$ and $90$, for our examples.

\mysection{\texorpdfstring{$\theta$}{theta}-Acceptability}
For service providers, acceptance is an important metric to plan, dimension and operate their services. Therefore, we would like to establish a link between opinion measurements
from subjective QoE studies and behavioral measurements. In particular, it would be very useful to derive the ``accept'' behavioral measure from opinion measurements of existing QoE studies. This would allow to reinterpret existing QoE studies from a business oriented
perspective. Therefore, we introduce the notion of $\theta$-acceptability which is based on opinion scores.

The $\theta$-acceptability, $\mathbb{A}_{\theta}$, is defined as the probability that the opinion score is above a certain threshold $\theta$, $P(U \ge \theta)$, and can be estimated by $\hat{f}_s$ from Eq.~\eqref{appendix:eq:p} or by counting all user ratings $U_i \geq \theta$ out of the $R$ ratings.
\begin{equation}\label{appendix:eq:accept}
\mathbb{A}_{\theta} = \int_{s = \theta}^{U^+} \hat{f}_s ds = \frac{1}{R} \left|\{U_i \geq \theta: i = 1, \dots, R \} \right|
\end{equation}

\mysection{Acceptance}
When a subject is asked to rate the quality as either {\em acceptable} or {\em not acceptable}, this means that $U$ is {\em Bernoulli}-distributed. The quality ratings are then samples of $U_i \in \{0,1 \}$, where \eqv{accepted}{1} and \eqv{not accepted}{0}.  The probability of acceptance is then $f_u = P(U=u)$, $U\in \{0,1\}$, and can be estimated by~\eqref{appendix:eq:p} with $u=1$:
\begin{equation}
\hat{f}_1 = \frac{1}{R} \sum_{i=1}^{R} \delta_{U_i,1}
\end{equation}

\mysection{\gob, \pow, and \tme}

\begin{figure}[h]%
\centering%
\includegraphics[width=\picwidth]{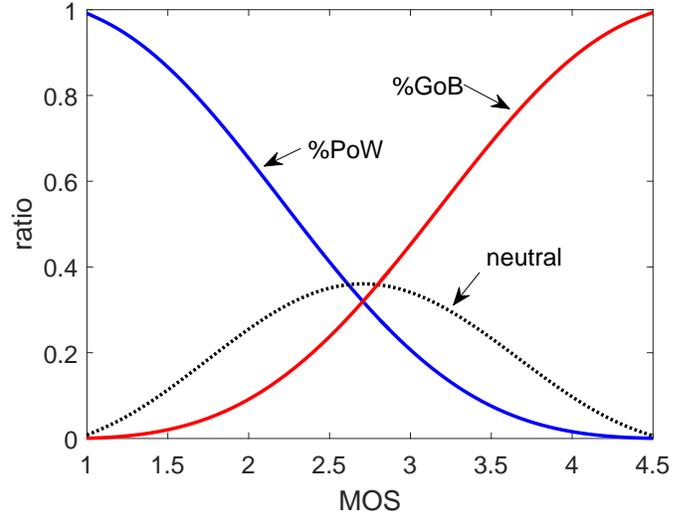}%
\caption{Relationship between MOS, \pow and \gob as used in the E-model \cite{ETR250}. The ratio of users not rating poor or worse as well as good or better is referred to as 'neutral' and is computed by $1-\gob-\pow$.}
\label{fig:rModelCDFlines}%
\end{figure}

In the past, service providers have also based their planning on (estimated)
percentages of users judging a service as ``poor or worse'' (\pow), ``good
or better'' (\gob), or the percentage of users abandoning a service (Terminate
Early, \tme). These percentages have been calculated from MOS distributions
on the basis of large collections of subjective test data, or of customer
surveys. Whereas the original source data is proprietary in most cases, the
resulting distributions and transformation laws have been published in some
instances. One of the first service providers to do this was Bellcore
\cite{Suppl.3}, who provided transformation laws between an intermediate
variable, called the Transmission Rating $R$, and \pow, \gob and \tme.
These transformation were further extended to other customer behavior
predictions, like retrial (to use the service again) and complaints (to the
service provider). The Transmission Rating could further be linked to MOS
predictions, and in this way a link between MOS, \pow and \gob could be
established. The E-model, a parametric model for planning speech telephony
networks, took up this idea and slightly modified the Transmission Rating
calculation and the transformation rules between $R$ and MOS, see \cite{ETR250}.
The resulting links can be seen in Fig.~\ref{fig:rModelCDFlines}. Such links can
be used for estimating the percentage of dissatisfied users from the ratings of
a subjective laboratory test; there is, however, no guarantee that similar
numbers would be observed with the real service in the field. In addition, the
subjective data the links are based on mostly stem from the 1970-1980s;
establishing such links anew, and for new types of services, is thus highly
desirable.

The measures {\em Poor-or-Worse} (\pow), {\em Good-or-Better} (\gob), and {\em Terminate Early} (\tme), are all quantile levels in the distribution of the quality rating $U$, 
or in the empirical distribution of $\mathcal{U} = \{ U_i \}$.

In the E-model~\cite{ETR250} the RV of the quality rating, $U \in [0;100]$ (referred to as the Transmission Rating $R$), 
represents an objective (estimated) rating of the voice quality.  
It is assumed that $U \sim N(0,1)$, which is the standard normal distribution, with the cumulative distribution function, 
\begin{equation}\label{append:eq:map}
F_U(u) = P(U\le u) = \int_{-\infty}^{u} \frac{e^{-\frac{t^2}{2}}}{\sqrt{2 \pi }} dt
\end{equation}

Under this assumption, the measures have been defined as\footnote{When $U \sim N(0,1)$ then $F_U(u)=1-F_U(-u)$, which is applied for the GoB definition.}
\begin{eqnarray}\label{appendix:eq:metrics1}
\gobn(u) &=& F_U(\frac{u - 60}{16})  = P_U(U\ge 60)  \\ \label{appendix:eq:metrics2}
\pown(u) &=& F_U(\frac{45 - u}{16}) = P_U(U\le 45)  \\ \label{appendix:eq:metrics3}
\tmen(u) &=& F_U(\frac{36 - u}{16}) = P_U(U\le 36)
\end{eqnarray}
(the \% of the measures are obtained by multiplying each of the measures by $100$).

The E-model also defines a transformation of the $U$ to a continuous scale MOS$ \in [1;4.5]$, by: 
\begin{equation}\label{appendix:eq:map}
MOS(u) = 7 \cdot (u-60) \cdot (100-u) \cdot u \cdot 10^{-6} + 0.035 \cdot u+1  
\end{equation}

By use of this transformation, the three measures are plotted for (continuous) MOS ($\in[1;4.5]$) in Figures~\ref{fig:rModelCDFlines}.
Observe that the sum of \gob + \pow do not add up to 100\%, 
because the probability (denoted "neutral" in the figure), $P(40 < U < 60)$, is not included in either \pow or \gob.
The quantiles used (i.e. $36$, $45$ and $60$) for the three measures, and the assumed standard normal distribution, 
are chosen as a result of a large number of subjective tests conducted while developing the E-model~\cite{ETR250}. 
Table~\ref{tab:trans} includes the MOS and the Transmission Rating $R$, with their corresponding values of the \pow, \gob, and \tme measures.  Observe: all values of MOS on the ACR scale are included, even for MOS=5 where the Transmission Rating $R$ is not defined. 
Furthermore, the table contains the MOS values (real numbers) for the quantiles for each of the three measures, i.e., when $R = 36, 45, 60$. 
\begin{table}%
\centering
\caption{MOS and Transmission Rating $R$ with the quantile measures.}
\label{tab:trans}
\begin{tabular}{rrrrr}
\toprule
MOS & $R$ & \pow & \gob & \tme \\
\midrule
 1.00000 &    6.52 &  99.192 &  0.041 &  96.732 \\
 1.50000 &  27.27 &  86.611 &  2.039 &  70.736 \\
\textcolor{red}{1.87293} &  \textcolor{red}{36.00} &  \textcolor{red}{71.311} &  \textcolor{red}{6.681} &  \textcolor{red}{50.000}\\
 2.00000 &  38.68 &  65.349 &  9.139 &  43.340 \\ 
\textcolor{red}{2.31513} &  \textcolor{red}{45.00} &  \textcolor{red}{50.000} &  \textcolor{red}{17.425} &  \textcolor{red}{28.689} \\
 2.50000 &  48.57 &  41.176 &  23.747 &  21.608 \\ 
 3.00000 &  58.08 &  20.685 &  45.221 &  8.381 \\ 
\textcolor{red}{3.10000} &  \textcolor{red}{60.00} &  \textcolor{red}{17.425} &  \textcolor{red}{50.000} &  \textcolor{red}{6.681} \\
 3.50000 &  67.96 &  7.563 &  69.062 &  2.288 \\ 
 4.00000 &  79.37 &  1.585 &  88.699 &  0.336 \\ 
 4.50000 &100.00 &  0.029 &  99.379 &  0.003 \\
 5.00000 & undefined & 0.000 & 100.000 & 0.000 \\
\bottomrule
\end{tabular}
\end{table}

The measures are estimated based on the ordered set of quality ratings, $\mathcal{U} = \{ U^{(i)} \}$, by using the $\theta$-Acceptability estimator from Eq.~\eqref{appendix:eq:accept}.
First, discretise the quality rating scale $\mathcal{U} \in \{0,100\}$. 
Then, using the Eq.~\eqref{appendix:eq:accept}, the following applies
\begin{eqnarray}\label{appendix:eq:ACR1}
\hat{\gob} &=&  \mathbb{A}_{\theta_{gb}} \\ \label{appendix:eq:ACR2}
\hat{\pow} &=& 1-\mathbb{A}_{\theta_{pw}} \\ \label{appendix:eq:ACR3}
\hat{\tme } &=& 1-\mathbb{A}_{\theta_{te}}
\end{eqnarray}
For example, in the E-model the $\theta_{gb}=60$, $\theta_{pw}=45$, and $\theta_{te}=36$ 
for $\mathcal{U} \in \{0,100\}$, and $\theta_{gb}=3.1$, $\theta_{pw}=2.3$, and $\theta_{pw}=1.9$ on a $\mathcal{U} \in \{1,5\}$ scale (when using Eq.~\ref{append:eq:map}).
\begin{eqnarray}
\hat{\gob } &=&  \mathbb{A}_{\lceil 3.1 \rceil } = \mathbb{A}_{4} = P(U\ge 4) \label{appendix:eq:gob2} \\ 
\hat{\pow } &=& 1-\mathbb{A}_{\lfloor 2.31513 \rfloor } =1-\mathbb{A}_{2}  = P(U\le 2)  \label{appendix:eq:pow2} \\ 
\hat{\tme } &=& 1-\mathbb{A}_{\lfloor 1.87293 \rfloor } = 1-\mathbb{A}_{1}  = P(U\le 1) \label{appendix:eq:tme2}
\end{eqnarray}

It is important to note that the quantiles in the examples are valid for {\em speech quality tests} under the assumptions given in the E-model. The mapping of MOS to the \pow and \gob metrics in Table~\ref{tab:trans} are specific for this E-model, but the \pow and \gob metrics are general and can be obtained from any quality study, provided that the thresholds $\theta_{gb}$ and $\theta_{pw}$ are determined.

The purpose of the example above is to demonstrate GoB, PoW, TME using an ACR scale (1-5). This is a theoretical exercise (valid for the E-model) where we apply the transformation from R to "MOS" (term used when E-model was introduced) as given in Eq.~\eqref{appendix:eq:map}, and transform Eq.~\eqref{appendix:eq:metrics1}-\eqref{appendix:eq:metrics3} into Eq.~\eqref{appendix:eq:ACR1}-\eqref{appendix:eq:ACR3}, using the notation introduced in Table~\ref{tab:trans}. Samples from Eq.~\eqref{appendix:eq:map} are given in Table~\ref{tab:trans}.
The $\gob = P(R \ge 60)$ corresponds to $\gob = P(\text{MOS} \ge 3.1)$ which on an integer scale is $\gob = P(\text{MOS} \ge 4)$.  Correspondingly, for $\pow = P(R \leq 45) = P(\text{MOS} \leq 2.32) = P(\text{MOS} \leq 2)$.  

\hfill

\newpage
\section{Details about Computation of SOS Parameter}

\subsection{Invariance of SOS Parameter for Linearly Transformed User Ratings}\label{appendix:sec:invarianceSOS}

In a subjective experiment, we observe the random variable (RV) $U$ for quality ratings (for a certain, fixed test condition). 
In the experiment, a continuous rating scale is used with lower bound $L_1$ and higher bound $H_1$, i.e. 
$U\in [L_1; H_1]$. We observe the SOS parameter $a$.

The expected value of $U$ is $E[U]=u$ with variance $Var[U]=V_1(u)$ according to the SOS hypothesis with 
\begin{equation}
V_1(u)=a_1(-u^2+(L_1+H_1)u-L_1\cdot H_1).
\end{equation}

Now, the user ratings $U$ are linearly transformed to another rating scale $[L_2;H_2]$ by the transformation function 
\begin{equation}
\tau(U) = \frac{U-L_1}{H_1-L_1}(H_2-L_2)+L_2.
\end{equation}

The variance of the transformed user ratings is 
\begin{eqnarray}
Var[\tau(U)]&=&Var\left[\frac{U-L_1}{H_1-L_1}(H_2-L_2)+L_2\right]\\
&=&Var\left[\frac{H_2-L_2}{H_1-L_1}U\right]\\ 
&=&\left(\frac{H_2-L_2}{H_1-L_1}\right)^2\cdot Var[U] .
\end{eqnarray}

However, the latter term is equivalent to the variance according to the SOS hypothesis with SOS parameter $a$ on the transformed rating scale, i.e.
\begin{equation}
V_2(\tau(u)) = a_2(-\tau(u)^2+(L_2+H_2)\tau(u)-L_2\cdot H_2) \, .
\end{equation}

For the user ratings transformed on the second rating scale, it holds 
\begin{equation}
V_2(\tau(u)) = Var[\tau(U)]
\end{equation}
which leads to 
\begin{equation}
a_2=a_1 = a \, .
\end{equation}

As a result, the SOS hypothesis holds with the same SOS parameter $a$. The SOS parameter $a$ is scale invariant when linearly transforming the user ratings in a mathematical way. However, it has to be clearly noted that subjective studies using different rating scales may lead to different SOS parameters. This has been observed e.g. for the results for speech QoE in~\cite{hossfeld2016qoeBeyond}.

\begin{figure}[th]%
\centering%
\includegraphics[width=\picwidth]{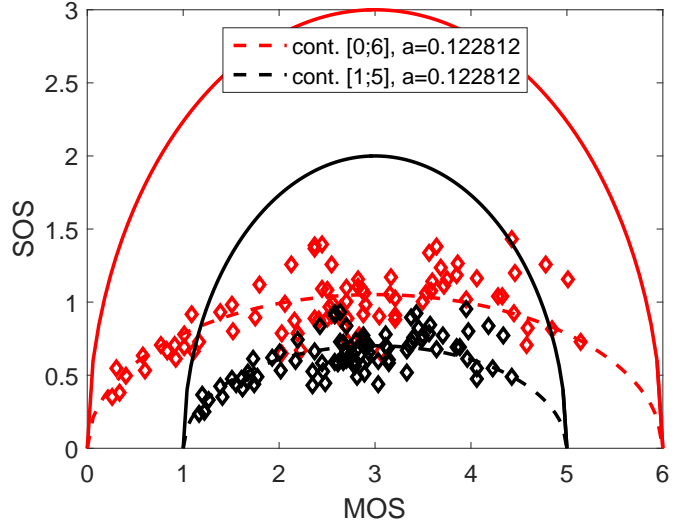}%
\caption{Transformation of user ratings from speech QoE results from rating scale $[0;6]$ to rating scale $[1;5]$ leads to the same SOS parameter $a$. However, the values of MOS and SOS (tuple depicted as diamond marker) as well as the maximum SOS for a given MOS (solid lines) are changing of course. 
}%
\label{fig:sosTransformed}%
\end{figure}

As an implication, the numerical derivation (by solving the optimization problem) of the SOS parameter $a$ for given MOS and SOS values can be done with linearly transformed user ratings, see Figure~\ref{fig:sosTransformed}. Thus, the SOS parameter reflects the user rating diversity independent of the rating scale.

\subsection{Computation of SOS Parameter}\label{sec:computationSOS}
Due to the scale invariance of the SOS parameter, we consider now normalized user ratings 
$Z \in [0;1]$ with $U^-=0$ and $U^+=1$. 
\begin{equation}
Z = \frac{U-U^-}{U^+-U^-}
\label{eq:normalizedScores}
\end{equation}
Then, the SOS hypothesis follows as
\begin{equation}
S(z) = \sqrt{a(-z^2 + z)}
\label{eq:computeSOS}
\end{equation}
for a normalized MOS value $z \in [0;1]$ with the SOS parameter $a$.

For deriving the SOS parameter $a \in [0;1]$ from the subjective normalized user ratings, for each test condition 
$c$ the MOS $u_c$ and the variance $\sigma^2_c$ are computed. Then, the least-squared error $L(a)$ between the subjective results and the SOS hypothesis is computed. The parameter $a$ is derived by deriving the minimal least squared error. 

\begin{align}
& L(a) = \sum_{c=1}^R \left( S(u_c)^2-\sigma^2_c \right) ^2 \\
 & \frac{d}{da} L(a) = 0 \\
\Rightarrow & a = - \frac{\sum_{c=1}^R(u_c^2-u_c)\sigma^2_c}{\sum_{c=1}^R(u_c^2-u_c)^2} 
\label{eq:computeSOSparameter}
\end{align}

\subsection{SOS Parameter on a Binary Scale}
We consider a binary scale where users accept ('1') or reject ('0') the observed quality. 
For any test condition $c$, we obtain a ration $p_c$ of users accepting the service quality ($U=1$). Hence, $1-p_c$ reject the service quality ($U=0$).

Thus, we observe the following mean and standard deviation for any %
test condition $c$.
\begin{eqnarray}
E[U] &=& u_c = p_c \label{eq:binaryMOS}\\
Var[U] &=& p_c(1-p_c) \label{eq:binaryVar}
\label{eq:acceptance}
\end{eqnarray}

According to the SOS hypothesis and due to the scale invariance, we observe the following relationship between a MOS value 
$u_c$ and ratio $p_c$ for any binary scale.
\begin{eqnarray}
	 S(u_c) &\underset{\text{Eq.(\ref{eq:computeSOS})}}{=}& \sqrt{a(u_c^2+u_c)} \underset{\text{Eq.(\ref{eq:binaryMOS})}}{=} \sqrt{a(-p_c^2+p_c)}  \\ 
		 S(u_c) &= & \sqrt{Var[U]} \underset{\text{Eq.(\ref{eq:binaryVar})}}{=} \sqrt{-p_c^2+p_c} \\
		\Rightarrow  & a = 1 &
\end{eqnarray}
Therefore, we conclude $a=1$. Using a binary scale (e.g. to ask about acceptance) leads to maximum user rating diversity. 

\section*{Acknowledgements}
This work was partly funded by Deutsche Forschungsgemeinschaft (DFG) under
grants HO 4770/1-2 and TR257/31-2 and in the framework of the COST ACROSS
Action. Mart{\'\i}n Varela's work was partially funded by Tekes, the Finnish
agency for research innovation, in the context of the CELTIC+ project NOTTS. The
authors alone are responsible for the content.

\end{document}